\documentclass[12pt]{iopart}

\usepackage{epsfig}
\usepackage{epstopdf}
\usepackage{epsf}
\input{epsf.sty}
\usepackage{graphicx}

\usepackage{iopams}

\newcommand{\calD}{{\cal D}}
\newcommand{\calG}{{\cal G}}
\newcommand{\calH}{{\cal H}}
\newcommand{\umax}{u_{\rm max}}
\newcommand{\xmax}{x_{\rm max}}
\newcommand{\ymax}{y_{\rm max}}
\def\be{\begin{equation}}
\def\ee{\end{equation}}
\def\bg{\begin{eqnarray}}
\def\nd{\end{eqnarray}}

\begin{document}

\title[]{A holographic model for large $N$ thermal QCD}

\author{Mohammed Mia$^{a}$, Keshav Dasgupta$^b$, Charles Gale$^{b}$ and Sangyong Jeon$^b$ }

\address{$^a$ Department of Physics,
Columbia University, 538 West 120th Street, New York, 10027, USA.\\
$^b$ Department of Physics, McGill University, 3600 University
Street, Montr{\'e}al QC, Canada H3A 2T8}
\ead{mm3994@columbia.edu; keshav,gale,jeon@hep.physics.mcgill.ca}
\begin{abstract}
We summarize the dual gravity description for a thermal gauge theory, reviewing
the key features of our holographic model of large $N$ QCD
\cite{FEP,jpsi1} and elaborating on some new results.
The theory has matter in the fundamental representation and the  gauge coupling runs logarithmically with
energy scale at low energies. At
the highest energies the theory becomes approximately scale invariant, much like what we would expect for
large $N$ QCD although not with asymptotic freedom. In this limit
the theory has a gravity dual captured by an almost classical supergravity description with a controlled quantum
behavior, such that
by renormalizing the supergravity action, we can
compute the stress tensor of the
dual gauge theory. From the stress tensor we obtain the shear viscosity and the entropy of the medium at a
temperature $T$, and 
the violation of the bound for the viscosity to the entropy ratio is then investigated.
By considering dynamics of open strings in curved spacetime
described by the supergravity limit, we compute the drag and diffusion coefficients for a heavy parton traversing the thermal medium. It is shown 
that both coefficients have a logarithmic dependence on momentum, consistent with pQCD expectations.
Finally, we study  the confinement/deconfinement
mechanism for quarks by analyzing open strings in the presence of the flavor seven branes.
We find linear confinement of quarks at
low temperatures,
while at high temperatures the quarkonium states melt, a behavior consistent  with the existence of a deconfined phase.

\end{abstract}

\maketitle

\section{Introduction}

A decade of running RHIC, the Relativistic Heavy Ion Collider at Brookhaven National Laboratory on Long Island, 
has produced  a plethora of intriguing experimental results that have challenged our understanding of hot and dense 
strongly interacting systems. One of these remarkable findings is that the matter produced at RHIC appears to display 
striking flow characteristics that are consistent with expectations from ideal relativistic hydrodynamics \cite{rhic_hydro}.  
In spite of the fact that the era of heavy ion physics at the LHC has barely begun, it appears that those exciting features 
also are present at higher energies \cite{schenke}. The commonly  accepted theoretical interpretation of this body of data is 
that the quark-gluon plasma (QGP) formed in the relativistic nuclear collisions performed at RHIC and the LHC is ``strongly coupled'': 
it displays aspects at odds with what perturbative QCD would predict.

Strongly coupled quark-gluon plasma (sQGP) poses theoretically challenging yet experimentally accessible questions. The
formation of QGP at RHIC is an example where theoretical descriptions are completely lacking at low energies
because our
perturbative techniques fail at strong couplings.  However
probing the non-perturbative  regime of large $N$
gauge theory through a {\it gravity} dual has led to
some interesting results for the physics of the quark-gluon plasma.
A popular approach so far has been the AdS/CFT correspondence \cite{Mal-1,Witt-1},
even though QCD is not a conformal field theory.
However, for certain gauge theories with running couplings,
there exist gravity duals. At zero temperatures these gravity duals are studied in \cite{O} with,
and in \cite{KS} without, fundamental flavors. On the other hand,
at high temperatures, there are examples of gravity duals with or without fundamental flavors in the literature
\cite{Karsh-Katz,Sakai,ahabu,cotrone}, all of which are quite successful in analyzing various aspects of strongly coupled nuclear matter in the IR. But
each of the models have their limitations, most prominently due to the absence of UV completions, and a supergravity description for
strongly coupled QCD incorporating various phases of quark
matter is yet to be discovered.

While the low energy limit of Klebanov-Strassler model with fundamental matter (which we call the
Ouyang-Klebanov-Strassler or the OKS model)
gives the most promising candidate to study
large $N$
strongly coupled QCD, the effective number of degrees of freedom grows indefinitely in the UV. In the presence of
fundamental flavors such a proliferation of degrees of freedom lead to Landau poles and UV divergences of
the Wilson loops.
However, since this cascading theory
provides an ideal setup to study phase transitions as one can describe various degrees of freedom in a single
framework, controlling the UV behavior then becomes essential to describe a consistent theory that could
describe physics at all energies.
In \cite{jpsi1} such a UV completion
was proposed and it is the goal of this paper to summarize and highlight some of the key features of this model and
discuss its relevance to
strongly coupled thermal QCD.

\section{Gauge theory and the Dual  Geometry}
Our goal is to construct the dual gravity of thermal field theory resembling features of strongly coupled QCD.
A gauge theory which becomes almost conformal in the UV with no Landau poles or UV divergences of the Wilson loops,
but has logarithmic running of
coupling in the IR, can be thought of as a suitable candidate mimicking QCD.
Following the developments in \cite{KS} and \cite{O},
in \cite{FEP} it was proposed that such a gauge theory indeed has a dual gravity. In fact the original proposal in
\cite{FEP} and \cite{jpsi1} was more along the {\it reverse} direction, namely, the IR geometry of \cite{O, KS}
was modified in steps in \cite{jpsi1} to construct a dual gauge theory that may have the required properties to be in the
same class as large $N$ QCD. On the gauge theory side we need IR confining and UV conformal. So the question is what
should we do to the gravity dual at large $r$, i.e at large radial coordinate,
to achieve the required properties?

The answer, as discussed in details in \cite{jpsi1}, is to first switch off the three-form fluxes at large $r$ so as
to keep only five-form fluxes. This way AdS type solutions could be constructed. Secondly, to get rid of the Landau
poles, we need to control the axio-dilaton behavior so that they decay rapidly at large $r$ instead of having their
usual growing logarithmic behavior.
Both these were achieved by switching on anti D5-branes and anti seven-branes with electric
and magnetic fluxes on them to kill off the unnecessary tachyons and thereby restoring stability.
Additionally the anti seven-branes, alongwith the seven-branes, were embedded in a
non-trivial way so as not to spoil the small $r$, or the IR, behavior of the theory \cite{jpsi1}. The anti D5-branes
are dissolved on the seven-branes, but they contribute to the three-form fluxes. Together they control both
the three form
fluxes as well as the axio-dilaton at large $r$.

To be a bit more precise,
the knowledge of the NS flux $B_2$ and dilaton $\phi$ uniquely determines the running of the gauge couplings.
On the other hand the
dual geometry is described by  the modified conifold \cite{jpsi1} and it is sourced by three form flux
$dB_2=H_{NS}, H_{RR}$,
five form flux
and $\overline{\rm D5}$-D7 branes. In the UV, $\overline{\rm D5}$-branes source three form fluxes
that exactly cancel the three form fluxes {\it originally}
sourced by the wrapped D5 branes \cite{jpsi1}. This way the effective three form
fluxes
 vanish
resulting in $B_2=0$. Then we can write
$g_1=g_2=g_{YM}$ at large $r$
and we get $g_{YM}\sim e^{\phi/2}$ $-$ thus the dilaton uniquely determines the gauge coupling. If
$\tau=C_0+ie^{-\phi}$ is the axio-dilation field, then using F-theory \cite{vafa} 
one can show  that $\tau$ behaves like an inverse power series for
large $r$, i.e $\tau\sim \frac{b_i}{r^i}, i>0$ \cite{jpsi1}, provided we have controlled the logarithmic behavior
by anti-brane embeddings discussed above.
In a more mathematical language of F-theory, $\tau$  is determined by finding the roots of the discriminant polynomial
$\triangle(z)$ where $z$ is the complexified coordinate
comprised of the two real coordinates of an internal 
${\mathbb P}^1$.
The roots give the precise location of the seven-branes and
hence the dilaton is uniquely determined by the seven-brane embeddings. Using the inverse power series behavior of $\tau$, and identifying the energy
scale with radial coordinate $r\sim \Lambda$,
one gets for the gauge coupling $g_1=g_2=g_{YM}$ the following behavior at the scale $\Lambda$:
\bg
g_{YM}\sim \frac{A_i}{\Lambda^i}
\nd
which indicates that the field theory reaches conformal fixed point at the highest energies.

Motivated by the above construction in \cite{FEP, jpsi1}, the question now would be how to
study the gauge theory, or the brane side of the story. The IR physics is captured by the near horizon
geometry of D3 branes at the conifold point with additional wrapped D5-branes. Thus in this picture one needs to
add the seven-branes as well as the anti-branes in addition to modifying the geometry. As we know,
any brane configurations that are
{\it away} from the original Klebanov-Strassler configuration, i.e not coinciding with the D3 and wrapped D5-branes,
would survive on the dual gravity side as branes (and not convert to pure geometry and fluxes). 
On the other hand, the modifications
that we did on the gravity side to get to the required gauge theory may be too difficult to simulate using simple 
arrangements of branes. However we can still incorporate many of the informations on the gravity side to the 
brane side of the story, that could at least give us a hint of the required gauge theory.  
Thus a possible proposal would be
that the following configuration of D3-D7-D5-branes (including the required anti-branes plus world-volume fluxes)
would give rise to our desired gauge theory, albeit in the large $N$
limit with matter in fundamental representation:
\begin{figure}[ht]
\begin{center}
\includegraphics[width=0.7\textwidth]{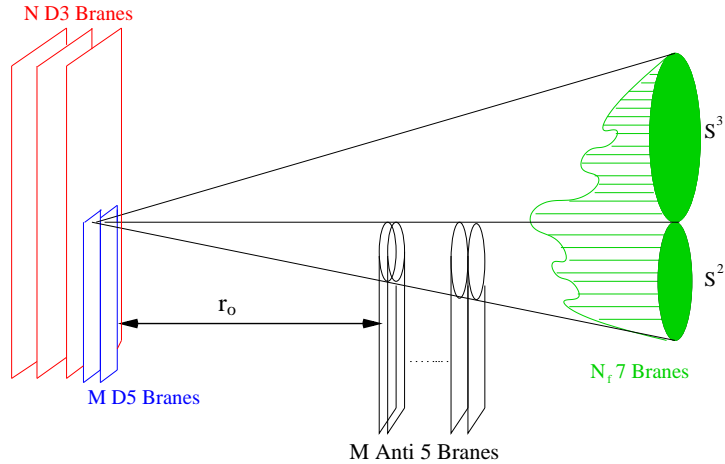}
\caption[]{A very simplified brane configuration in ten dimensional geometry ${\cal M}_4\times {\cal C}_6$, 
where ${\cal M}_4$ is Minkowski space and ${\cal
C}_6$ is a six dimensional cone. The actual configuration is non-trivial to depict pictorially and may not 
even exist as a simple arrangement of branes. The gravity dual, on the other hand, is more robust and can be 
easily described as shown in \cite{jpsi1}.}
\label{charm_xsec}
\end{center}
\end{figure}
Place $N$ D3 branes at the tip of
a six dimensional conifold with radial coordinate $r$ and base $T^{1,1}=S^2\times S^3$. The world volume of the D3 branes
extend in four Minkowski directions. Now place $M$ number of D5 branes which wrap the $S^2$ on the conifold base and
extend
in four Minkowski directions and place another $M$ number of anti {D}5 branes away from but parallel to the D5 branes.
The D5-$\overline{\rm D5}$ branes are separated along the radial $r$ direction. Furthermore
 place $N_f$ number of D7 branes extending in the radial $r$ direction,
wrapping  the base $T^{1,1}$
 and filling up four
Minkowski directions. The brane - anti brane configurations will result in tachyonic modes but introducing fluxes sourced
by the branes can
stabilize the system. Details of the embedding can be found in \cite{jpsi1}, \cite{O} and one can, with the risk of
oversimplification and possible pitfalls, depict the brane configuration as in Fig. 1.

The excitations of these D3-D5-D7 branes (and the required anti-branes)
are described by a gauge theory which in the UV has  $SU(N+M)\times
 SU(N+M)$
color symmetry group and $SU(N_f)\times SU(N_f)$ flavor symmetry group \cite{jpsi1} from the
seven-branes. This is because addition of ($p, q$) branes at the junction, or more appropriately
anti five-branes at the junction with gauge fluxes on its world-volume, tell us that the number of three-branes
degrees of freedom are $N + M$, with the $M$ factor coming from five-branes anti-five-branes pairs.
Furthermore, the $SU(N + M) \times
SU(N + M)$ gauge theory will tell us that the gravity dual is approximately AdS, but has RG flows because of the
fundamental flavors.
At the scale $r = r_0$ we expect one of the gauge group
to be Higgsed, so that we are left with $SU(N + M) \times SU(N)$. Now both the gauge fields flow at different rates
and give rise to the cascade that is slowed down  by the $N_f$ flavors. In the end, at far IR, we expect
confinement at zero temperature.
If $g_1, g_2$ are the two gauge
couplings, they are related to
the dilaton $\phi$ and the two form NS-NS flux $B_2$ by following duality relations that capture the essence of
gauge/gravity duality here:
\bg \label{coupling1}
\frac{4\pi^2}{g_1^2}+\frac{4\pi^2}{g_2^2}&=&\frac{\pi}{e^{\phi}}\nonumber\\
\left[\frac{4\pi^2}{g_1^2}-\frac{4\pi^2}{g_2^2}\right]e^{\phi}&=&\frac{1}{2\pi\alpha'}\int B_2
\nd
A little bit of elaboration may elucidate the story further.
As the anti five branes are placed at some radial location $r_0$, their excitations give massive modes with energy
scale $\Lambda_0\sim r_0$. At low energies, $\Lambda < \Lambda_0$, these modes are not excited and hence can be ignored.
This means at
low energies, the effect of the anti five branes $\overline{\rm D5}$ will not be important and we effectively have
 Ouyang-Klebanov-Strassler geometry of \cite{FEP} with a black-hole, which is region 1 of \cite{jpsi1}.
 Thus in the IR we will have the $SU(N+M)
 \times SU(N)$  color symmetry group and $SU(N_f)\times SU(N_f)$ flavor symmetry group with
gauge couplings $g_1, g_2$ running
 logarithmically as $B_2 \sim M{\rm log}~r$. More precisely, using the exact form of fluxes and the 
dilaton at small $r$, i.e at low energies,
 we get the following beta functions:
\bg \label{beta_1}
\frac{\partial}{\partial{\rm log}~\Lambda}\left[\frac{4\pi^2}{g_1^2} + \frac{4\pi^2}{g_2^2}\right] &=&
- \frac{3N_f}{4}\nonumber\\
\frac{\partial}{\partial{\rm log}~\Lambda}\left[\frac{4\pi^2}{g_1^2} - \frac{4\pi^2}{g_2^2}\right] &=&
3M\left(1 + \frac{3g_s N_f}{2\pi} ~{\rm log}\Lambda\right)
\nd
At the IR, we see that the two gauge couplings run in opposite directions and $SU(N+M)$ flows to strong
coupling. Performing a Seiberg duality
transformation, we identify the strongly coupled $SU(N+M)$ with a weakly coupled $SU(N-(M-N_f))$ at the IR.
We see that not only the number of colors are reduced, but the difference of the size of the gauge group now decreases
from $M$ in \cite{KS}
to $M-N_f$. This difference will decrease by the increments of $N_f$ until it is smaller than
or equal to $N_f$. Then there are two possible end points:

\vskip.1in

\noindent $\bullet$ If $N$ is still greater than zero then we will have an
approximately conformal theory in the far IR, or

\vskip.1in

\noindent $\bullet$ If $N$ decreases to zero but with finite $M$ left over then we will have a
$SU(M)$ theory with $N_f$ flavors that confines in the far IR (see
\cite{O} for more details).

\vskip.1in

\noindent The latter theory, or
more particularly the high temperature limit of the latter theory, is what we are interested in and henceforth we will
only consider that theory\footnote{For a review of Seiberg dualities and cascading theories,
see \cite{strasslerreview}. For a review on brane constructions for cascading theories see \cite{DM}.}.
Once we know the dual background we can say that the
Hilbert space of the gauge theory can be obtained from the Hilbert space of the string theory on this
geometry. The dual geometry of above brane theory in the decoupling regime was proposed in  \cite{jpsi1}. One basically minimizes
 SUGRA action with the fluxes running logarithmically with $r$ for small $r$ and as inverse of $r$ for large $r$.
The resulting metric in the non extremal
limit is
\bg \label{metric1}
ds^2&=&-\frac{g(r)}{\sqrt{h}}dt^2+
\frac{H\sqrt{h}}{g(r)}dr^2
+\frac{1}{\sqrt{h}}d\overrightarrow{x}^2
+r^2\widetilde{g}_{mn}dx^m dx^n\nonumber\\
&\equiv& {g}_{\alpha\beta} dx^\alpha dx^\beta
+r^2\widetilde{g}_{mn}dx^m dx^n = G_{\mu\nu}dx^\mu dx^\nu
\nd
where $\alpha,\beta=0,1,2,3,4$, $m,n=5,6,7,8,9$ and $\mu,\nu=0,..,9$. The internal five-dimensional
metric is denoted by $\widetilde{g}_{mn}$ whereas the other five-dimensional space is denoted by ${g}_{\alpha\beta}$.
We have also defined the following:
\bg \label{metric2}
h \equiv h(r,\psi,\phi_1,\phi_2,\theta_1,\theta_2)&= &\frac{L^4}{r^4}\left[1+\sum_{i=1}\frac{a_i(r_h,\psi,\theta_1,\theta_2)}{r^i}\right]~~~ {\rm  for ~~ large} ~r\\
&= & \frac{L^4}{r^4}\left[\sum_{j,k=0}\frac{b_{jk}(r_h,\psi,\theta_1,\theta_2){\rm log}^k(r)}{r^j}\right]~~~
{\rm  for ~~ small} ~r\nonumber
\nd
for the warp factor $h$. Note that $h$ is taken to be functions of all the internal as well as
the radial coordinate $r$. This should be compared to earlier work of \cite{KS} where the
warp factor is only a function of the radial coordinate. The rest of the variables are now defined as:
\bg
 H \equiv H(r,\psi,\phi_1,\phi_2,\theta_1,\theta_2)& = & \sum_{i=0}\frac{c_i(r_h,\psi,\theta_1,\theta_2)}{r^i}~~~ {\rm  for ~~ large} ~r\\
& = & \sum_{j,k=0}\frac{d_{jk}(r_h,\psi,\theta_1,\theta_2){\rm log}^k(r)}{r^j}~~~
{\rm  for ~~ small} ~r\nonumber
\nd
%POPO
with $g(r)= 1-\frac{r_h^4}{r^4}$,
where $r_h$ is the black-hole horizon and $a_i,b_{jk},c_i,d_{jk}$ are  determined by the location of the D7 branes, the fluxes and the
horizon. As discussed in \cite{Herzog-Gubser}, the non extremal solution has a regular Schwarzchild horizon that covers the IR singularity of
the extremal metric but the internal space gets
modified. This modification of the internal space is described by the metric $\widetilde{g}_{mn}$ which is not the metric of $T^{1,1}$ but
accounts for the deformation of the internal space.
The forms of $a_i,b_{jk},c_i,d_{jk}$ and $\widetilde{g}_{mn}$ are given in \cite{jpsi1} for the extremal case and the extremal solution of the IR
geometry is discussed in \cite{FEP}. The non extremal solution for the brane configuration in Fig 1 is under preparation \cite{nex-geom}.
However, the study of confinement mechanism and the computation of diffusion and drag coefficient will only require the knowledge of the warp
factors at the location of the string which we will choose to be fixed in the internal directions. This means
$(\psi,\phi_1,\phi_2,\theta_1,\theta_2)$ will be constant along the string and we will treat $a_i,b_{jk},c_i,d_{jk}$ as
only functions of horizon radius $r_h$.

To compute the stress tensor, we do need to know the exact form of $a_i,b_{jk},c_i,d_{jk}$ as a
function of internal coordinates. However, to obtain information about the four dimensional field theory, we will first integrate over the internal
directions and after the integrations, we will be left with a five dimensional metric $g_{\alpha\beta}$ of the form given in (\ref{metric1}) with warp factors $h,H$ now
only being functions of $r$. Then the coefficients ${a}_i, {b}_{jk}, {c}_i, {d}_{jk}$ of the five dimensional warp factors would
only depend on $r_h$.

The question now is
how does the warp factor $h$ depend on the black-hole horizon $r_h$? Of course the Einstein equations along with the flux equations will
dictate this dependence- but one can restrict the form of the warp factor as a function of $r_h$ by analyzing the dual gauge theory.
In particular, the warp factor is related to the size of the dual gauge group \cite{KS}\cite{FEP},
\bg \label{Neff1}
&&g_s N_{\rm eff} \alpha'=r^4h=L^4\left(1+\sum_{i=1}\frac{a_i}{r^i}\right)\nonumber\\
&&\Rightarrow N_{\rm eff}=N\left(1+\sum_{i=1}\frac{a_i}{r^i}\right)
\nd
at large $r$. Now from the T-dual description of the brane configuration in Fig 1 \cite{Thesis}, one concludes that $N_{\rm eff}$ keeps on growing with
energy scale $\Lambda$ i.e. the radial coordinate $r$ on the dual gravity side,
which is depicted by Fig 2. From 
%\eqref{Neff1}
\ref{Neff1}
we see that this means
\bg \label{cond11}
\sum_k {k a_k \over r^{k-1}} ~ < ~ 0
\nd
for large $r$. But $a_k$'s are functions of $r_h$, thus (\ref{cond11}) can be used to restrict the form $a_k$'s
as functions of $r_h$. For example one
way is that all $a_k(r_h) < 0$ for all $r_h$ so that this will automatically satisfy (\ref{cond11}).

On the other hand
for small $r$, one possibility is that the IR geometry is completely hidden behind the black-hole horizon,
which will be the case
for high temperatures \cite{Herzog-Gubser}. Then we are left with the geometry at large $r$.
However for low temperature and small $r$, we have
\bg \label{Neff}
&&g_s N_{\rm eff} \alpha'=r^4h=L^4\left[\sum_{j,k=0}\frac{b_{jk}(r_h,\psi,\theta_1,\theta_2){\rm log}^k(r)}{r^j}\right]\nonumber\\
&&\Rightarrow N_{\rm eff}=N\left[\sum_{j,k=0}\frac{b_{jk}(r_h,\psi,\theta_1,\theta_2){\rm log}^k(r)}{r^j}\right]
\nd
Again from the gauge theory interpretation that $N_{\rm eff}$ grows with $r$, we get
\bg \label{cond2}
\sum_{j, k} \left[\frac{j b_{jk}~{\rm log}^k(r)}{r^{j-1}} ~ - ~   \frac{k b_{jk}~{\rm log}^{k-1}(r)}{r^{j+1}} \right] ~
 < ~ 0
\nd
 As long as (\ref{cond11}) is satisfied at large $r$ and (\ref{cond2}) is satisfied at small $r$, we have a consistent supergravity description
 for thermal cascading gauge theory which flows according to Fig 2.
\begin{figure}[ht]
\begin{center}
\includegraphics[width=0.45\textwidth]{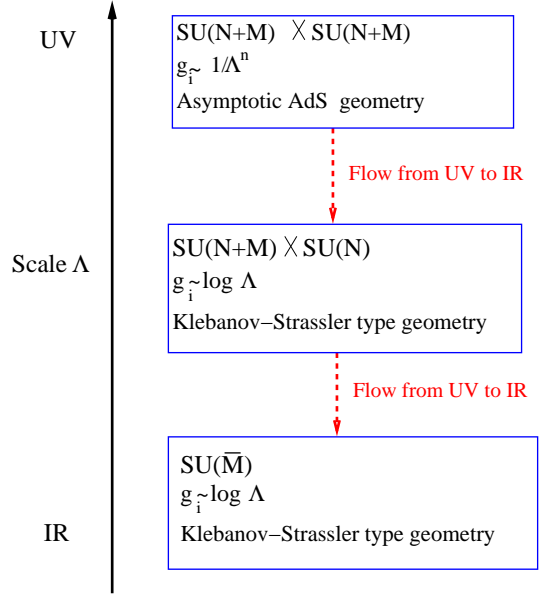}
\caption[]{The flow in gauge theory and its dual gravity description. A cascade of Seiberg dualities reduces the rank of the gauge group as
energy scale is altered from UV to IR.}
\label{charm_xsec}
\end{center}
\end{figure}

%popo

\section{The Stress Tensor and $\eta/s$}
\label{bla3}
In order to compute expectation values of certain observables  of the gauge theory, one first computes the partition
function. As the Hilbert space of the quantum field theory arising from brane excitations in Fig 1 is contained in the Hilbert space of
the geometry sourced by these branes, there should be a one-to-one correspondence between the partition functions of the gauge
theory and the dual gravity. A precise mapping for the conformal case
was proposed by Witten \cite{Witt-1}  who stated that the partition function of
the strongly coupled quantum field theory should be identified with the partition function of the weakly coupled classical
gravity.

In particular if we are interested in computing the expectation value of an operator $\langle{\cal O}\rangle$
with a source
$\phi_0$, one can make the following identification of the partition function as a functional of the source $\phi_0$:
\bg \label{KS16}
{\cal Z}_{\rm gauge}[\phi_0] &~\equiv &~ \langle{\rm exp}\int_{M^4} \phi_0 {\cal O}\rangle
~=~ {\cal Z}_{\rm gravity}[\phi_0] \nonumber\\
&~ \equiv &  ~{\rm exp}(S_{\rm SUGRA}[\phi_0] + S_{\rm GH}[\phi_0] + S_{\rm counterterm}[\phi_0])
\nd
where $M^4$ is a Minkowski manifold, $S_{\rm GH}$ is the Gibbons-Hawking boundary term \cite{Gibbons-Hawking},
$\phi_0$ should be understood as a fluctuation over a given configuration of field
and $S_{\rm counterterm}$ is the
counter-term action added to renormalize the action. We will briefly describe how each term is obtained from supergravity
action and their necessity.

First observe that $\phi$ is a bulk field, which  is in  general a function of the
coordinates of  the
geometry and $\phi_0$ is the boundary value of $\phi$. One integrates the classical ten dimensional
gravity action over
the compact manifold ${\cal M}_5$ to obtain an effective action $S^{\rm eff}_5$ for a five dimensional manifold.
 The bulk five dimensional geometry has a boundary at $r=\infty$ and the boundary describes a four dimensional manifold with
an induced Minkowski flat metric. Thus the boundary value of
$\phi$
\bg
\phi_0(t,x,y,z)\equiv \phi(r=\infty,t,x,y,z)
\nd
is a field which lives in four dimensional flat space and plays the role of source in the four dimensional gauge theory.

Now to obtain the gravity action as a functional of $\phi_0$, one  integrates  the five dimensional effective action $S^{\rm
eff}_5$
over the  radial coordinate $r$ and we are left with gravity action $S[\phi_0]$
as a function of $\phi_0$. It turns out that in general for the
gravity action $S[\phi]$ to be stationary under perturbation $\phi\rightarrow
 \phi+\delta\phi$, one needs to add surface terms $S_{\rm GH}[\phi]$
which are known as the Gibbons-Hawking terms \cite{Gibbons-Hawking}. On the
 other hand, after the radial integral is done, for some sources $\phi$, $S[\phi_0]$ becomes infinite and one needs to
 regularize the action to obtain finite expectation values for operators. The regularization requires addition of extra
 terms which cancel the infinities that appear in the gravity action and are denoted by $S_{\rm counterterm}$.
 Keeping all this in mind, one takes the functional
 derivative of (\ref{KS16}) with respect to $\phi_0$ to obtain
 \bg \label{O}
\langle{\cal O}\rangle&=&\frac{\delta {\cal Z}_{\rm gauge}}{\delta \phi_0}=\frac{\delta {\cal Z}_{\rm gravity}}{\delta
\phi_0}\nonumber\\
&\sim&~{\rm exp}\left(S_{\rm total}[\phi_0]\right)~\frac{\delta S_{\rm total}[\phi_0]}{\delta\phi_0}\Big\vert_{\phi_0=0}
\nd
where we depict $S_{\rm total}[\phi_0]$ as:
$S_{\rm total}[\phi_0]=S[\phi_0] + S_{\rm GH}[\phi_0] + S_{\rm counterterm}[\phi_0]$. Note that, from our construction,
this is also the renormalized action.

Observe that in the usual AdS/CFT case we consider the action at the boundary to
map it directly to the dual gauge theory side. For general gauge/gravity dualities, there are many
possibilities of defining
different gauge theories at the boundary depending on how we cut-off the geometry and add UV
caps. The details on adding geometries at large $r$ to existing IR geometries for small $r$
can be found in \cite{FEP,jpsi1}.

One of the important thermodynamical quantity to extract from our background would be the stress tensor.
As the stress tensor couples to the metric, its expectation value is obtained from the renormalized action
$S_{\rm total}$
via
$\langle T^{ij}(\Lambda_c)\rangle=\frac{\delta S_{total}[l_{ij}]}{\delta l_{ij}}\Big{|}_{l_{ij}=0}$
where we have chosen
$\phi_0 \equiv l_{ij}$ as the four dimensional effective metric induced by the ten dimensional metric of the
 form (\ref{metric1}) (see \cite{FEP} for details on the procedure).
At every fixed $r$ the five dimensional metric induces a four dimensional metric and with appropriate rescaling of the
time coordinate, we can view this effective
four dimensional metric as a flat Minkowski metric (with possible perturbations).
If $O_{ij}$ denote the perturbations on the
background by quark strings that stretch from the D7 branes to the horizon of the black hole,
then writing the supergravity action up to quadratic order in $O_{ij}$; using integration by
parts together with appropriate Gibbons-Hawking terms; and holographically renormalizing the subsequent action, 
we obtain:
\bg\label{wakeup}
&& T^{mm}_{{\rm medium} + {\rm quark}}
 = \int \frac{d^4q}{(2\pi)^4}\sum_{\alpha, \beta}
\Bigg\{({H}_{\vert\alpha\vert}^{mn}+ {H}_{\vert\alpha\vert}^{nm})s_{nn}^{(4)[\beta]}
-4({K}_{\vert\alpha\vert}^{mn}+ {K}_{\vert\alpha\vert}^{nm})s_{nn}^{(4)[\beta]}\nonumber\\
&& ~~~~~~~~~ +({K}_{\vert\alpha\vert}^{mn}+ {K}_{\vert\alpha\vert}^{nm})s_{nn}^{(5)[\beta]}
+\sum_{j=0}^{\infty}~\hat{b}^{(\alpha)}_{n(j)} \widetilde{J}^n
\delta_{nm}  e^{-j{\cal N}_{\rm uv}} + {\cal O}(T e^{-{\cal N}_{uv}})\Bigg\}
\nd
where ($\hat{b}^{(\alpha)}_{n(j)},
{\cal N}_{\rm uv}$) together will specify the full boundary theory for a specific
UV complete 
theory\footnote{It may appear from 
(\ref{wakeup})
above that the energy-momentum tensor is {\it diagonal}. This is 
however not so because we are using the variables $L_{mm}$, defined in eq. (3.108) of \cite{FEP}, to write 
(\ref{wakeup}).
However if we replace $L_{mm}$ by $l_{\mu\nu}$, defined in eq. (3.107) of \cite{FEP}, the energy-momentum tensor 
will not appear diagonal. We simply found it convenient in \cite{FEP} to express the energy-momentum tensor using the 
$L_{mm}$ variables. These subtleties are described in full details in \cite{FEP}.}.
The quantities $H^{mn},\widetilde{J}^{m},K^{mn}$ etc. with
$m,n=0,..,3$ are given in \cite{FEP, jpsi1} and are
completely independent of the radial coordinate $r$ and depend on the 3+1 dimensional spacetime
coordinates. Note also that the stress tensor depends only on the temperature $T$ and is independent of any cut-offs. The
dependence on the UV degrees of freedom is exponentially suppressed, and in the limit
${\cal N}_{\rm uv} = \epsilon^{-n}, n >> 1$
we reproduce the result for the parent cascading theory.

This exponential suppression comes from the form of  $N_{\rm eff}=r^4 h$ where $h$ is the warp factor. For geometries
such that $N_{\rm eff}(r)\sim N(1+{\rm log}~r)$ at large $r$, one easily gets that $r^j\sim e^{j N_{\rm eff}(r)}=e^{j {\cal N}_{\rm uv}}$
where we have taken $r\rightarrow \infty$ limit and defined $N_{\rm eff}(\infty)={\cal N}_{\rm uv}$. Sending $r\rightarrow \infty$ also means
${\cal N}_{\rm uv}\rightarrow \infty$, thus the final result for the stress tensor will be same for various UV completions. However, the point
we are trying to make is that different UV completions reach the same IR limit.

For Asymptotic AdS spaces, we have the warp factors behaving like (\ref{metric2}) and $N_{\rm eff}(\infty)={\cal  N}_{uv}=N(1+a_0)$. Then for the stress tensor we get
\bg\label{wakeup1}
&& T^{mm}_{{\rm medium} + {\rm quark}}
 = \int \frac{d^4q}{(2\pi)^4}\sum_{\alpha, \beta}
\Bigg\{({H}_{\vert\alpha\vert}^{mn}+ {H}_{\vert\alpha\vert}^{nm})s_{nn}^{(4)[\beta]}
-4({K}_{\vert\alpha\vert}^{mn}+ {K}_{\vert\alpha\vert}^{nm})s_{nn}^{(4)[\beta]}\nonumber\\
&& ~~~~~~~~~~~ +({K}_{\vert\alpha\vert}^{mn}+ {K}_{\vert\alpha\vert}^{nm})s_{nn}^{(5)[\beta]}
+\hat{a}^{(\alpha)}_{n} \widetilde{J}^n
\Bigg\}
\nd
where $\hat{a}^{(\alpha)}_{n}$ is uniquely determined by $a_0$ which appears in (\ref{Neff}) and thus determines ${\cal N}_{uv}$. Here we have
already set $r^{-j}$ terms to zero, so they don't appear in (\ref{wakeup1}), and dependence on ${\cal N}_{uv}$ appears implicitly through
$\hat{a}^{(\alpha)}_{n}$. More precisely, for a generic asymptotic AdS UV completion, there might not be any analytic inverse functions ${\cal
F}$ such
that $r\sim {\cal F}({\cal N}_{uv})$. Hence it might not be possible to see how $r^{-j}\sim {\cal F}^{-j}({\cal N}_{uv})$ reaches zero. But
the final result is independent of how this limit is reached and is only sensitive to  $\hat{a}^{(\alpha)}_{n}$, thus (\ref{wakeup1}) uniquely
determines the field theory stress tensor.

With the
general formulation of stress tensor (\ref{wakeup}),(\ref{wakeup1}),
which is similar to the AdS/CFT results \cite{Skenderis} in the limiting case $M= N_f=0$,
we can compute the wake a moving quark creates in a plasma with $O_{ij}$ being the
metric perturbation due to string (see \cite{Yaffe} for an equivalent AdS calculation).
Furthermore we can compute the shear viscosity $\eta$ from the Kubo formula
with the propagator obtained from the dual partition function (\ref{KS16}) and the entropy density
$s$ obtained by computing energy and pressure density using the stress tensors (\ref{wakeup}), (\ref{wakeup1}). Alternatively entropy $s$ can
be identified with the black hole entropy using Wald's formula and
the result for the ratio $\eta/s$ is given by:
\bg \label{final1}
\frac{\eta}{s} &=&~ {1 + \sum_{k = 1}^\infty
\alpha_k e^{-4k {\cal N}_{\rm uv}} \over 4\pi + {1\over \pi}~{\rm log}^2 \left(1 -
{T}^4 e^{-4{\cal N}_{uv}}\right)}\\
&-&\frac{c_3\kappa}{3 L^2 \left(1- {T}^4 e^{-4{\cal N}_{uv}}\right)^{3/2}}
 \left[\frac{{B_o}(4\pi^2-{\rm log}^2 ~C_o)+4\pi{A_o}~{\rm log}~C_o}{\Big(4\pi^2-{\rm
log}^2~C_o\Big)^2+16\pi^2~{\rm
log}^2~C_o}\right] ~ < ~ {1\over 4\pi}\nonumber
\nd
where ($A_o, B_o, C_o, \alpha_k$) are constants that depend on the temperature $T$ and $e^{-{\cal N}_{\rm uv}}$; and
$c_3$ is the coefficient of the Riemann square term coming from the back reactions of the D7 branes in the background.
These have been explicitly worked out in \cite{FEP}. The $c_3$ dependence of the $\eta/s$ ratio first appeared
in \cite{kats} and the above expression shows that for $c_3>0$, we have violation of the lower bound for $\eta/s$. For $c_3$ greater than some
critical value, studies have shown that the theory becomes acausal \cite{etasv1}-\cite{etasv7}. This restricts the form of the terms that
appears from D7 brane back reaction as we require the boundary gauge theory to be causal.  Observe that for asymptotic AdS space,
using (\ref{wakeup1}) for stress tensor, $\eta/s$  is given by (\ref{final1})
with ${\cal N}_{uv}=\infty$ and then the only violation comes from the Riemann square term.

\section{Momentum broadening and drag}

Consider a parton moving through a plasma in four dimensional Minkowski spacetime with the following world line
\bg \label{wl1}
x(t)&=&vt+\delta x(t)\nonumber\\
z(t)&=&y(t)\equiv \delta y(t)
\nd
For a fast moving parton, we can always choose coordinates such that (\ref{wl1}) is the world line. Now if the plasma has
matter in fundamental representation, has logarithmic running of coupling in the IR that becomes asymptotically conformal,
we can treat the ten dimensional geometry with metric (\ref{metric1}) to be the
dual gravity of this gauge theory which lives in four dimensional flat spacetime.

To obtain the momentum broadening of the
parton  we shall use the Wigner distribution function $f$ as defined in QCD kinetic theory \cite{KT}
\bg
f(X,r_\perp)&\equiv& \langle f_{cc}(X,r_\perp) \rangle ~
=\rm{Tr}\left[\rho Q_a^{\dagger}(X_-^\perp)U_{ab}(X_-^\perp,X_+^\perp)Q_a(X_+^\perp)\right]\nonumber\\
f(X,r_L)&\equiv& \langle f_{cc}(X,r_L)\rangle~
=\rm{Tr}\left[\rho Q_a^{\dagger}(X_-^L)U_{ab}(X_-^L,X_+^L)Q_a(X_+^L)\right]
\nd
where $X_-^\perp=X-r_\perp/2$,$X_+^\perp=X+r_\perp/2$,$X_-^L=X-r_L/2$, $X_+^L=X+r_L/2$,
$X=(t,\bf{x})$ is the world line of the parton field $Q_a$ without any fluctuation, $U_{ab}$ is the link and $\rho$ is
the density matrix. The index $ab$ refers to color and for our choice of the world line for the parton we have
$r_\perp^2=2\delta y^2,r_L^2=\delta x^2$. Now Fourier transforming the distribution functions, we can get  the average
transverse momentum to be
\bg
\langle p^2_\perp \rangle ~ =\int d^3x\int\frac{d^2k_\perp}{(2\pi)^2}k^2_\perp f(X, k_\perp)
\nd
whereas the average longitudinal momentum {fluctuation} is given by
\bg
\langle p^2_L \rangle ~=\int d^3x\int\frac{d^2k_L}{(2\pi)^2}k^2_L f(X, k_L)
\nd
Now if we assume that the initial transverse momentum and longitudinal momentum fluctuation distributions are
narrow, then we
have
\bg
\langle p^2_\perp \rangle~ &=&2\kappa_T {\cal T}\nonumber\\
\langle p^2_L \rangle~ &=&\kappa_L {\cal T}
\nd
with ${\cal T}$ some large time
interval, and $\kappa_T, \kappa_L $ are the diffusion coefficients. Generalizing the arguments in \cite{DTCS} to include
longitudinal momentum fluctuations,  one can write the diffusion coefficients solely in terms of functional derivative of
Wilson loops.  The final result is
\bg \label{kappaTL}
\kappa_T&=&\lim_{\omega\rightarrow 0}\frac{1}{4}
\int dt \;e^{i\omega t}\;\left[iG_{11}^y(t, 0)+iG_{22}^y(t, 0)+iG_{12}^y(t, 0)+iG_{21}^y(t, 0)\right]\nonumber\\
\kappa_L&=&\lim_{\omega\rightarrow 0}\frac{1}{4}
\int dt \;e^{i\omega t}\;\left[iG_{11}^x(t, 0)+iG_{22}^x(t, 0)+iG_{12}^x(t, 0)+iG_{21}^x(t, 0)\right]
\nd
 where the Green's functions are
\bg \label{prop}
G_{11}^\zeta(t,t')&=& \frac{1}{\rm{tr}\rho^0 W_C[0,0]}\Bigg\langle \rm{tr}\rho^0 \frac{\delta^2 W_C[\delta \zeta_1,0]}{\delta
\zeta_1(t)\delta
\zeta_1(t')}\Bigg\rangle\nonumber\\
G_{22}^\zeta(t,t')&=& \frac{1}{\rm{tr}\rho^0 W_C[0,0]}\Bigg\langle \rm{tr}\rho^0
\frac{\delta^2 W_C[0,\delta \zeta_2]}{\delta \zeta_2(t)\delta
\zeta_2(t')}\Bigg\rangle\nonumber\\
G_{12}^\zeta(t,t')&=& \frac{1}{\rm{tr}\rho^0 W_C[0,0]}\Bigg\langle \rm{tr}\rho^0 \frac{\delta^2 W_C[\delta \zeta_1,\delta \zeta_2]}{\delta
\zeta_1(t)\delta
\zeta_2(t')}\Bigg\rangle\nonumber\\
G_{21}^\zeta(t,t')&=& \frac{1}{\rm{tr}\rho^0 W_C[0,0]}\Bigg\langle \rm{tr}\rho^0 \frac{\delta^2
W_C[\zeta_2(t),\zeta_2(t')]}{\delta \zeta_1(t)\delta
\zeta_2(t')}\Bigg\rangle
\nd
 with $\zeta=y,x$ and  $t,t'$ are the real part of complex time $t_C,t_C'$ on contour $C$ (fig 3).
  Here we have introduced type `1' and `2' fields ($\delta \zeta_i,i=1,2$) of thermal field theory
 evaluated along the upper and lower horizontal line of
the contour $C$.
 We denote by $W_C[\delta \zeta_1,\delta \zeta_2]$ the Wilson loop with deformation $\delta \zeta_1$ and  $\delta \zeta_2$ on
 the upper and lower lines of $C$.
\begin{figure}[ht]
\begin{center}
\includegraphics[width=\textwidth]{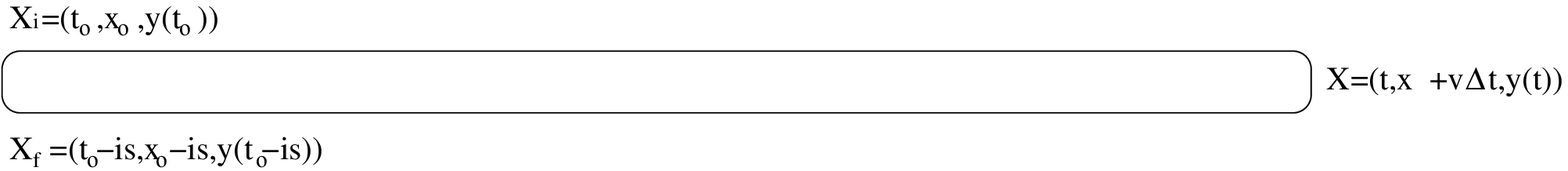}
\caption[]{The contour C with $X$ denoting the coordinates of a point on the contour. The upper line with real time coordinate is the
worldline of the heavy parton while the lower line has complex time coordinate.
Fields evaluated on the upper line are the type `1' fields while those evaluated
at lower line are type `2' fields of thermal field theory. Here $s\rightarrow 0$ is real.}
\label{charm_xsec}
\end{center}
\end{figure}

The Wilson loop at strong coupling can be computed  using holography. This means we identify
\bg
\langle {\rm tr}\rho^0W_C \rangle ~ =e^{iS_{\rm NG}}
\nd
where $\rho^0$ is the density matrix \cite{DTCS}, $S_{\rm NG}$ being the Nambu-Goto action, with the boundary
of the string worldsheet being  the curve $C$. That is the string worldsheet ends on the world line (\ref{wl1}) of the  heavy parton in four
dimensional flat spacetime.
If $X^\mu:(\sigma, \tau)\rightarrow
(t,x, y, z,\zeta\equiv 1/r,\psi,\phi_1,\phi_2,\theta_1,\theta_2)$ is a mapping from string worldsheet to ten-dimensional geometry,
 we restrict to the case when
\bg\label{lcone} \nonumber
&&X^0=t,~~X^4=r,~~X^1=x(\tau,\sigma),~~X^2=X^3=~\delta y(\tau,\sigma)\nonumber\\
&& X^k=0 ~(k=5,6,7),~~(X^8, X^9) = (\theta_1, \theta_2) =\pi
\nd
 With the choice of parameterization\footnote{We denote the axio-dilaton and the world-sheet time using the same symbol
$\tau$ (as is the usual lore in string theory). It should be clear from the context which one is meant.}
$\tau=t,\sigma=\zeta$ also known as the static gauge,
we have
\bg \label{pert1}
x(t, r)=vt+\bar{x}(r)+\delta x(t,r)\nonumber\\
z(t,r)=y(t, r)=\delta y(t,r)
\nd
where $\bar{x}$ is the unperturbed solution for the mapping. Minimizing the Nambu-Goto action gives the fields $x(t,r),y(t,r)$ and
$z(t,r)$ and
 finally integrating $S_{\rm NG}$ over radial direction $r$, one gets the boundary action. We can obtain the Green's function from the boundary action and the result
 for the diffusion coefficients are \cite{Mia_d}
\begin{eqnarray} \label{kappaT}
\kappa_T&=&\pi \sqrt{\gamma g_sN_{\rm eff}} T^3 (1+{\cal B})\nonumber\\
\kappa_L&=&\gamma^{5/2}\pi \sqrt{g_sN_{\rm eff}} T^3 (1+{\cal B})
\end{eqnarray}
where $\gamma=1/\sqrt{1-v^2}$, ${\cal B}$ is a constant of ${\cal O}(g_sM^2/N, g_s^2M^2N_f/N)$ which depends on 
the velocity $v$, and is defined in the following way:
\bg\label{BB}
{\cal B}&=&\frac{\frac{a}{2}\;{\rm log}(\gamma)-b \;{\rm log}(r_h) {\rm log}(\gamma)+\frac{b}{4}{\rm log}^2(\gamma)}
{1-a\;{\rm log}(r_h)
+b\;{\rm log}^2(r_h)}\nonumber\\
a&=&\frac{3g_s M^2}{2\pi N}+\frac{3g_s^2 M^2N_f}{4\pi^2 N}\nonumber\\
b&=&\frac{9g_s^2M^2N_f}{4\pi^2N}
\nd
Here $N_{\rm eff}$ is the number of effective degrees of freedom for the boundary
gauge theory
\begin{eqnarray} \label{NEff}
N_{\rm eff}&=&N\Bigg[1+\frac{27g_s^2M^2N_f}{32\pi^2 N}-\frac{3g_sM^2}{4\pi N}+\left(\frac{3g_sM^2}{4\pi
N}-\frac{9g_s^2M^2N_f}{16\pi^2 N}\right){\rm log}\left(\frac{r_c}{r_0}\right)\nonumber\\
&+&\frac{9g_s^2M^2N_f}{8\pi^2N}{\rm log}^2 \left(\frac{r_c}{r_0}\right)\Bigg]
\end{eqnarray}
where $r_c$ is the radial distance where the warp factor changes from a logarithm to a power series and $r_0$ is associated with the maximum
depth of the seven branes in the radial direction. The value of
$r_c$ and $r_0$ will of course be determined by the embedding equations for the branes which give their precise location in ten dimensional
space. Thus for different configuration of branes in
ten dimensions, we will end up with different effective degrees of freedom for the gauge theory in four-dimensional
Minkowski space. Note for duality to hold, $N$ must be quite large, making $N_{\rm eff}$ quite large. In deriving (\ref{NEff}) above, we utilized the logarithmic behavior of the warp
factors $h,H$ for $r\sim r_c$ and the precise form for $h$ given by,
\bg \label{warph}
h&=&\frac{L^4}{r^4}
\Bigg[1-\frac{3g_sM^2}{2\pi N}{\rm log}\left(\frac{r_0}{r}\right)\left\{1+\frac{3g_sN_f}{2\pi}\left(-{\rm log}\left(\frac{r_0}{r}\right) +\frac{1}{2}\right)
\right\}\Bigg]
\nd
For $N_f=M=0$, we get
back the value of $\kappa_T,\kappa_L$ as computed in \cite{DTCS}\cite{Gubser}. However for a non conformal field theory with fundamental matter
- which is more relevant for QCD - $M\neq 0, N_f \neq 0$, and our analysis shows that the diffusion coefficients $\kappa_T$ and $\kappa_L$ depends non
trivially on velocity $v$ of the parton. In fact $\kappa_T,\kappa_L$ varies as $\rm{log}(1-v^2)$ and increases with velocity. This dependence on
velocity is consistent with perturbative calculation of diffusion for heavy partons due to collisions with quarks and gluons using Langevin
dynamics \cite{Guy-Teaney}.  Details of our calculation can be found in \cite{Mia_d}.

Now we compute the drag experienced by a parton as it ploughs through a thermal plasma at constant speed using dual gravity description where
 the parton is represented by a string moving in ten dimensional geometry with metric (\ref{metric1}). We consider the map (\ref{lcone}),
 (\ref{pert1}) in the limit $\delta x=\delta y=0$. From the Nambu-Goto action one derives the following
canonical momenta for the string:
 \bg
 && \Pi_\mu^0= -T_0 G_{\mu\nu}\frac{(\dot{X}\cdot X')(X^{\nu})'-(X')^2(\dot{X}^\nu)}{\sqrt{\cal K}}\nonumber\\
&& \Pi_\mu^1 = -T_0 G_{\mu\nu}\frac{(\dot{X}\cdot X')(\dot{X}^{\nu})-(\dot{X})^2(X^\nu)'}{\sqrt{\cal K}}
 \nd
 where $T_0$ is the string tension $1/2\pi \alpha'$, $G_{ij}$ is the metric given by (\ref{metric1}), ${\cal K}$ is the determinant of the
 world sheet without any fluctuations, prime denotes derivative with respect to $r$ while dot denotes derivative with
 respect to $t$. Now the rate at which momentum is lost to the black hole is given by the momentum density at horizon
\bg \label{KS15a}
\Pi_1^x(r=r_h)= - T_0 Cv
\nd
while the force  the parton experiences due to friction with the  plasma is $\frac{dp}{dt}=-\nu p$ with
$p=mv/\sqrt{1-v^2}$. To keep the parton moving at
constant velocity, an external field ${\cal E}_i$ does work and the equivalent energy is dumped into the
medium \cite{HK1}\cite{HK2}. Thus
the rate at which a quark dumps energy and momentum into the thermal medium is
precisely the rate at which the string loses energy and momentum to the black hole. Thus up to
${\cal O}(g_sN_f, g_s M)$ we have ${\nu m
v \over \sqrt{1-v^2}} = - \Pi_1^x(r=r_h)$ and
\bg \label{KS15b}
\nu &= & \frac{ C \sqrt{1-v^2}}{2\pi \alpha' m}\\
&=&\frac{1}{2\pi \alpha' mL^2}\frac{r_h^2}
{\sqrt{1+\frac{3 g_sM^2}{2\pi N}~{\rm log}\Big[\frac{r_h}{(1-v^2)^{{1}/{4}}}\Big]
\Big(1+\frac{3 g_sN_f}{2\pi}~\Big\{{\rm
log}~\Big[\frac{r_h}{(1-v^2)^{{1}/{4}}}\Big]+\frac{1}{2}\Big\}\Big)}}\nonumber\\
&\equiv& \frac{T^2}{2\pi \alpha' m} \left(1+{\cal C}\right)
\nd
which defines ${\cal C}$ and it depends on velocity $v$. More precisely, ${\cal C}\sim {\rm log}(\gamma)$, similar to ${\cal B}$ given in
(\ref{BB}).

In determining $\nu$, we used the logarithmic behavior for the warp factor $h$ near horizon $r=r_h$ given by (\ref{warph}).
The above result for drag should now be compared with the AdS result \cite{HK1}\cite{HK2}. In the AdS limit, $M=N_f=0$ and we get back the AdS drag.
However, unlike the AdS case where drag is independent of velocity, the result
in (\ref{KS15b}) is velocity dependent. In particular it decreases with increasing velocity- a result consistent with the perturbative
analysis of \cite{Guy-Teaney}.

\section{Confinement/deconfinement from holography}

 The linear confinement of quarks at large separations and low
temperatures is a strong coupling phenomenon and one uses lattice QCD  to compute the free energy of the bound state of
quarks \cite{Wilson-1,Susskind,Polyakov}. On the other hand at high enough temperatures, if the gauge coupling is weak, one can compute the free energy using
perturbative QCD and obtains Coulombic interactions between the quarks.
Whether using perturbative QCD at weak coupling or
lattice QCD and effective field theory techniques at strong couplings, one finds that at high temperatures, the Coulomb
potential is Debye screened  \cite{Gale-kapustaVqq}\cite{Susskind}.

The study of heavy quark potential gathers special attention as it is linked to a possible signal from Quark Gluon
Plasma. In particular, heavy quarks are formed during early stages of a heavy ion collision as only then there exists enough
 energy for their formation. On the other hand right after the collision  QGP is formed as only then the temperature is
 high enough for quarks to be deconfined.  As temperature goes down, heavy quarks form bound states and as these states are
 very massive,
they are formed at temperatures higher than deconfinement value.  This  means heavy quark bound states like  $J/\psi$ can
coexist with QGP and can act as probes to the medium.

In particular one can study the $J/\psi$ bound state formed in
proton-proton collisions and compare with heavy ion collision where a medium is formed. The medium will screen the
$c\bar{c}$ interaction making them less bound and eventually resulting in a suppression  of $J/\psi$ production. This
phenomenon is known as the $J/\psi$ suppression and considered as a signal of QGP formation \cite{Matsui-Satz-1}.

In order to quantify quarkonium suppression and analyze features of plasma that causes the suppression, one has to compute
free energy of $J/\psi$. At large couplings, lattice QCD calculations are most reliable
and there has been extensive studies quantifying the potential for heavy quarks \cite{Fkarsch-1}-\cite{HQG-4} .
The effective potential between the quark anti-quark pairs separated by a distance $d$
at temperature ${T}$
can then be expressed
succinctly in terms of the free energy $F(d, {T})$, which generically takes the following form:
\bg\label{freeenergy}
F(d,  T) = \sigma d ~f_s(d, T) - {\alpha\over d} f_c(d, T)
\nd
where $\sigma$ is the string tension, $\alpha$ is the gauge coupling and $f_c$ and $f_s$ are the screening
functions\footnote{We expect the screening functions $f_s, f_c$ to equal identity when the temperature goes to
zero. This gives the zero temperature Cornell potential.}
(see for example \cite{Fkarsch-1}-\cite{Fkarsch-5} and references therein). At zero temperature free energy is the potential energy of the pair.
On the other hand free energy or potential energy of quarkonium is related to the Wilson loop which we will elaborate here.

Consider the Wilson loop of a rectangular path ${\cal C} $ with space like width $d$ and time like length ${\cal T}$.
Note that we are working with Euclidean metric with time $\tau=it$ and the time like
paths can be thought of as world lines of pair of quarks $Q\bar{Q}$ separated by a spatial distance $d$.
Studying the expectation value of the Wilson loop in the
limit ${\cal T}\rightarrow \infty$, one can show that it behaves as
\bg \label{WL-1}
\langle W({\cal C})\rangle ~\sim~ {\rm exp}(-{\cal T} E_{Q\bar{Q}})
\nd
where $E_{Q\bar{Q}}$ is the energy of the $Q\bar{Q}$ pair which we can identify with their
potential energy $V_{Q\bar{Q}}(d)$ as the quarks are static. At this
point we can use the principle of holography \cite{Mal-1} \cite{Mal-2}
and identify the expectation value of the Wilson loop with
the exponential of the {\it renormalized} Nambu-Goto action,
\bg \label{Holo-1}
 \langle W({\cal C})\rangle ~\sim ~ {\rm  exp}(-S^{\rm ren}_{{\rm NG}})
\nd
with the understanding that ${\cal C}$ is now the boundary of string world sheet.
Note that we are computing Wilson loop of gauge theory living on
flat four dimensional space-time $x^{0, 1, 2, 3}$. Whereas the string world sheet is embedded in curved five-dimensional
manifold with coordinates
$x^{0, 1, 2, 3}$ and $r$. We will identify the five-dimensional manifold with Region 3 of \cite{jpsi1}. For the
correspondence in (\ref{Holo-1}) to be valid, we need the t'Hooft  coupling which is the gauge coupling in the theory to be
large. On the other hand as discussed before, it is in this regime of strong coupling that linear confinement is realized in
gauge theories. Thus using gauge/gravity duality is most appropriate in computing quarkonium potential. Comparing (\ref{WL-1}) and (\ref{Holo-1})
we can read off the
potential
\bg \label{Vqq}
V_{Q\bar{Q}}~ = ~ \lim_{{\cal T} \to \infty} \frac{S^{\rm ren}_{{\rm NG}}}{{\cal T}}
\nd
Thus knowing the renormalized
string world sheet action, we can compute $V_{Q\bar{Q}}$ for a strongly coupled gauge theory.

For non-zero temperature, the free energy is related to  the Wilson lines $W\left(\pm {d\over 2}\right)$
 via:
\bg\label{wlfe}
{\rm exp}\left[-{F(d, T)\over T}\right] ~ = ~
{\langle W^\dagger\left(+{d\over 2}\right) W\left(- {d\over 2}\right)\rangle \over
\langle W^\dagger\left(+{d\over 2}\right)\rangle \langle W\left(-{d\over 2}\right)\rangle}
\nd
In terms of Wilson loop, the free energy (\ref{freeenergy}) is now related to the renormalized Nambu-Goto
action for the string on a background with a black-hole\footnote{There is a big literature on the
subject where quark anti-quark potential has been computed using various different approaches like
pNRQCD \cite{HQG-1}$-$\cite{HQG-4}, hard wall AdS/CFT \cite{polstrass}$-$\cite{boschi4} and other
techniques \cite{reyyee}$-$ \cite{cotrone2}. Its reassuring
to note that the results that we get using our newly constructed background matches very well with the results
presented in the above references. This tells us that despite the large $N$ nature there is an underlying
universal behavior of the confining potential.}. One may also note that the theory we get is a four-dimensional theory
{\it compactified} on a circle in Euclideanised version and not a three-dimensional theory.

We will restrict to  cascading gauge theories where the effective 
number of colors grows as the scale grows. This property of a
gauge theory is most relevant for physical theories as new degrees of freedom emerge at 
UV and effective degrees of freedom
shrink in IR to form condensates at low energy \cite{strasslerreview}.
The number of colors at any scale $u = 1/r$ is given by (\ref{Neff})
and for the analysis given here, it is simpler to define
\be
\calH(u) \equiv {u^2\over \sqrt{h}} =
{\sqrt{N}\over L^2\sqrt{N_{\rm eff}}}
%=
%{1\over L^2\sqrt{1+ a_l u^l}}
\label{eq:F_of_u}
\ee
instead of $N_{\rm eff}(u)$.
In terms of $\calH(u)$, the condition that $N_{\rm eff}(u)$ is
a decreasing
function of $u=1/r$ becomes
\be
\calH'(u) > 0
\label{eq:fprime}
\ee
Combining (\ref{eq:F_of_u}) and (\ref{eq:fprime})
yields the following condition
\be
\calH(u) > {1\over L^2}
\ee
{}From now on, the value of $L$ is set to $1$ for the rest of this subsection, so that $\calH(u)> 1$.

\subsection{Zero temperature}

Let $\umax$ be the maximum value of $u$ for the string between
the quark and the anti-quark.
Then the relationship between $\umax$ and
the distance between the quark and the anti-quark is given by \cite{jpsi1}

\bg \label{d}
d(\umax) =
2\umax
\calH(\umax)
\int_{\epsilon_0}^1dv\,
{v^2 \sqrt{\calG_m \umax^m v^m}
\over (\calH(\umax v))^2}
\left[1-v^4\left(\calH(\umax)\over \calH(\umax v)\right)^2\right]^{-1/2}
\label{eq:dumax}
\nd
while the renormalized Nambu-Goto action can be written as \cite{jpsi1}
\bg \label{NG-4}
S_{\rm NG}^{\rm ren}&=&\frac{{\cal T}}{\pi}\frac{1}{u_{\rm max}}
\Bigg\{-{\widetilde{\cal G}}_0+ \sum_{l=2}\frac{{\widetilde{\cal G}}_l}{l-1}
- \int_0^1 {dv\over v^2} \sqrt{{\cal G}_m u_{\rm max}^m v^m} +
{\cal O}(g_s^2) \nonumber\\
&+& \int_0^1 {dv\over v^2} \sqrt{{\cal G}_m u_{\rm max}^m v^m}
\left[1-v^4\left(\calH(\umax)\over \calH(\umax v)\right)^2\right]^{-1/2} + {\cal O}(\epsilon_o)\Bigg\}
\nd
where $\epsilon_0$ gives the depth of the seven brane on which the string is attached, ${\cal G}_m,\tilde{{\cal G}}_m$ depends on the metric
and is explicitely given in \cite{jpsi1} for our supergravity background.

Observe that for $\umax$ small, $d$ is small and ignoring higher order terms in $\umax$, we get from (\ref{d}) and (\ref{NG-4})
\bg\label{sdpot}
V_{Q\bar{Q}} ~&& = ~ -\left({a_0\vert b_0\vert\over \pi}\right) {1\over d} ~+~ \left({b_1\over \pi a_0}\right)d
+ {\cal O}(d^3)\nonumber\\
&&= ~ -{0.236\over d} ~+~\left(0.174{\cal G}_2 + 0.095{\cal A}_2\right) d ~ + ~ {\cal O}(d^3)
\nd
where ${\cal A}_2, {\cal G}_2$ depend on the geometry and their values for our metric (\ref{metric1}),(\ref{metric2}) will depend on the
coefficients $a_i,c_i$ which are computed in \cite{jpsi1}. However, the potential is dominated by the inverse $d$ behavior, i.e
the expected Coulombic behavior and the coefficient of the
Coulomb term which is a dimensionless number is independent of the warp factor and therefore should be universal.
This result, in appropriate units,
is of the same order of magnitude as the real Coulombic term obtained by comparing with  Charmonium
spectra as first modeled by \cite{charmonium} and subsequently by several authors \cite{Fkarsch-1}-\cite{Fkarsch-5},\cite{HQG-1}-\cite{HQG-4},\cite{boschi1}-\cite{boschi4}. This
prediction, along with the overall minus sign, should be regarded as a success of our model (see also \cite{zakahrov}
where somewhat similar results have been derived in a string theory inspired model). The second term on the other
hand is model dependent, and vanishes in the pure AdS background.

Now we will analyze (\ref{d}) and (\ref{NG-4}) for large $\umax$. First note that for (\ref{d}) to represents the physical
distance between a quark and an anti-quark in vacuum,
the integral must be real. This is guaranteed if for all   $0 \le v \le 1$:
\bg \label{cond1}
W(v|\umax) \equiv v^2\left( {\calH(\umax)\over \calH(\umax v)}\right) \le
1
\nd
For AdS space, (\ref{cond1}) is automatic, as ${\cal H}=1$ and then $d$ is proportional to $\umax$ which results in
only Coulomb potential for all values of $d$.
But for a generic warp factor,
(\ref{cond1}) gives rise to a finite upper bound for $\umax$ and we will analyze the behavior of $d$ and the action $S_{\rm NG}^{\rm ren}$ near the
upper bound.

To show confinement at large distances the potential between the quark and the
anti-quark
must be long ranged. That is, $d(\umax)$ must range from 0 to $\infty$ as
$\umax$ varies from 0 to it's upper bound, say $\umax=\xmax$.
Since $\calH(u)> 1$ for our non-AdS geometry, the only way to satisfy these conditions is via sufficiently fast vanishing of the square-root
in (\ref{eq:dumax}) as $v\to 1$ at
$\umax=\xmax$. Studying the behavior of the square-root \cite{jpsi1}, one concludes that
 $\xmax$ must be
the smallest positive solution of
\be
x\calH'(x) - 2\calH(x) = 0
\label{eq:zeroTcond}
\ee
For our dual geometry with metric (\ref{metric1}),(\ref{metric2}), indeed (\ref{eq:zeroTcond}) has a real positive solution $\xmax$.
As $\umax$ approaches $\xmax$, both $d$ and $S_{\rm NG}^{\rm ren}$ become infinite, growing linearly with $\umax$ and we get \cite{jpsi1}
\bg\label{lipo}
V_{Q\bar Q} ~ = ~ \left(\frac{{\cal H}(\xmax)}{ \pi \xmax^2}\right) ~d
\nd
i.e linear confinement of quarks separated by large distance at zero temperature.

\subsection{Finite temperature}

At finite temperature, the relation between $\umax$ and the distance
between
the quark and the anti-quark is obtained by
replacing $\calH(u)$ with $\sqrt{1-u^4/u_h^4}\,\calH(u)$ in
(\ref{eq:dumax}):
\bg
d_T(\umax)&=&2\umax
\sqrt{1-\umax^4/u_h^4}\calH(\umax)
\int_{\epsilon_0}^1dv\,
{v^2 \sqrt{\calD_m \umax^m v^m}
\over (1-v^4\umax^4/u_h^4)(\calH(\umax v))^2}\nonumber\\
&&\times
\left[
1-v^4{(1-\umax^4/u_h^4)\over(1-v^4\umax^4/u_h^4)}
\left(\calH(\umax)\over \calH(\umax v)\right)^2
\right]^{-1/2}
\label{eq:dumaxT}
\nd
The explicit factor of $\umax$ makes $d_T(\umax)$ vanish at $\umax=0$
as in the $T=0$ case.
As $\umax$ approaches $u_h$, the integral near $v=1$ behaves like
\bg
d_T(\umax)\sim
\int_0^1 dv
{\sqrt{1-\umax^4/u_h^4} \over \sqrt{(1-v)(1-v\umax/u_h)}}
\nd
which indicates that $d_T(\umax)$ goes to 0 as $\umax$ approaches $u_h$.
Hence, at both $\umax=0$ and $\umax=u_h$, $d_T(\umax)$ vanishes.
Since $d_T(\umax)$ is positive in general,
there has to be a maximum between $\umax=0$ and $\umax=u_h$.
Whether the maximum value of $d_T(\umax)$ is infinite as in the $T=0$ case
depends on the temperature (equivalently, $u_h^{-1}$) as we now show.

The fact that the physical distance needs to be real
yields the following condition. For all $0 \le v \le 1$,
\be
W_T(v|\umax)\equiv
v^2\left(\calH(\umax)\over \calH(\umax v)\right)
\sqrt{1-\umax^4/u_h^4\over 1-\umax^4 v^4/u_h^4} \le 1
\ee
Similarly to the $T=0$ case, $d_T(\umax)$ can have an infinite range
if  $\umax = \ymax$ is
the smallest positive solution of the following equation
\be
y{\bf \calH}'(y) - 2\calH(y) = (y/u_h)^4\,y\calH'(y)
\label{eq:finiteTopt}
\ee

Note that the left hand side is the same as the zero temperature
condition,
(\ref{eq:zeroTcond}). The right hand side is the temperature $(u_h)$
dependent part.
Using the facts that:
\bg\label{knownfact}
\prod_{k=0}^{j-1}\left(k-1/2\right)
= -(2j-3)!!/2^j,~~~ \sum_{j=1}^\infty x^j\, (2j-3)!!/2^j j!
= 1- \sqrt{1-x},
\nd
it is clear that $y=u_h$ cannot be a solution of
(\ref{eq:finiteTopt}) because at $y=u_h$, the equation reduces to
$\calH(u_h) = 0$ which is inconsistent with the fact that $\calH(y)\ge 1$.

Recall that we are considering gauge theories for which $\calH(y)\ge 1$ and $\calH'(y)\ge 0$, and thus the right hand side of
(\ref{eq:finiteTopt}) is always non-negative.
As $y$ increases from 0 towards $\xmax$,
the left hand side of (\ref{eq:finiteTopt})
increases from $-2$
while the right hand side increases from 0.
Assuming that the warp factors do not get corrections from black hole horizon, i.e. ${\cal H}(u,u_h=0)={\cal H}(u,u_h)$ for all $u_h$, we see
that left hand side reaches 0 when $y = \xmax$
which is the point where the distance
$d(\umax)$ at $T=0$ becomes infinite.
At this point the right hand side of (\ref{eq:finiteTopt}) is
positive and has the value
$(\xmax/u_h)^4\xmax\calH'(\xmax)$.
Hence the solution of (\ref{eq:finiteTopt}), if it exists,
must be larger than $\xmax$.

Even if we consider ${\cal H}$ to be a function of horizon $u_h$, we expect solution of (\ref{eq:finiteTopt})
$y=\ymax>\xmax$. This is because black holes will attract the U shaped string toward the horizon and a string attached to quarks separated
by a distance $d$ will stretch deeper into the bulk geometry in the presence of a black hole.  A warp factor $h(u,u_h)$ that
explicitely depends on horizon and thus ${\cal H}(u,u_h)$ explicitely depending on $u_h$ will not alter this attraction of the black hole
and we will have  $\ymax>\xmax$.

Consider first low enough temperatures so that $u_h \gg \xmax$.
For these low temperatures,
(\ref{eq:finiteTopt}) will have a solution, as the right hand side will be still small around $y =\xmax$.
This then implies that
the linear potential at low temperature will have
an infinite range if the zero temperature potential has an infinite range and we get
\bg\label{lipo}
V_{Q\bar Q} ~ = ~ \left(\frac{{\cal H}(\ymax)}{ \pi \ymax^2}\right) ~d
\nd

Now we  show that the infinite range potential cannot be maintained at all
temperatures. We can have a black hole such that $u_h = \xmax$.
When the left hand side vanishes at $y=\xmax$,
the right hand side is
$\xmax\calH'(\xmax) = 2\calH(\xmax)$
which is positive and finite.
For $y > \xmax$, the left hand side ($y\calH'(y)-2\calH(y)$)
may become positive, but it is always smaller than $y\calH'(y)$
since $\calH(y)$ is always positive.
But for the same $y$, the right hand side ($(y/u_h)^4y\calH'(y)$)
is always positive and necessarily larger than
$y\calH'(y)$ since $(y/u_h) > 1$.
Hence, (\ref{eq:finiteTopt}) cannot have a real and positive
solution when $u_h = \xmax$.
Therefore between $u_h = \infty$ and $u_h = \xmax$, there {\it must be a point $u_h=u_c$
when (\ref{eq:finiteTopt}) cease to have a positive solution}. Thus for some $u_h<u_c$ and hence for temperature $T>T_c$,
(\ref{eq:finiteTopt}) has no solution and
$d_T(\umax)$ will not diverge for any $\umax$
within $(0, u_h)$. Furthermore, since the expression vanishes at both
ends,
there must be a maximum $d_T(\umax)$ at a non-zero $\umax$.
When the distance between the quark and the anti-quark is greater than
this
maximum distance, there can no longer be a string connecting the quark and
the anti-quark. Hence for $T>T_c$ the string breaks and we have two parallel straight strings describing the quarks. Of course
there will be interactions between the two straight strings but these integrations will be $\alpha'$ suppressed and thus can be ignored in the
$\alpha'\rightarrow 0$ limit which is the decoupling limit where duality is precise. Hence we can treat these quarks described by straight
strings as `free' quarks.

To summarize, we have just shown that if we start with a dual geometry that allows infinite range linear potential at zero
temperature,
there exists some critical temperature above which the string connecting the quarks break. This shows that at high enough
temperatures quarkonium state melts and gives rise to `free quarks'.

Thus our dual gravity analysis indicates that if we have long range interactions of $Q\bar{Q}$ pairs at
zero temperature where the potential energy grows linearly with distance, this interaction {\it cannot} be sustained at higher temperatures.
This is the key point of our analysis, as this shows that quarks that are linearly confined at low temperatures, {\it must} be deconfined at
higher temperatures and thus we have a confinement/deconfinement phase transition.

In Fig 4 we plot of interquark separation $d$ as a function of $\umax$ while in Fig 5 we plot free energy denoted by $V_{Q\bar{Q}}$ as a function
of separation $d$. We have identified the renormalized Nambu-Goto action in the presence of a black hole with the free energy of $Q\bar{Q}$
pair at finite temperature i.e. $F(d,T)\sim S_{\rm NG}^{\rm ren}/{\cal T}$ where ${\cal T}=\int d\tau$ with $\tau$ being Euclidean time. This
way of obtaining free energy is indeed consistent with the identification of partition functions of gauge theory with that of gravity.
For the plots in Fig 4 and Fig 5, we have made a particular choice for the warp factors
\bg\label{warpy}
h~=~ L^4 u^4 {\rm exp}(-\alpha u^{\widetilde{\alpha}}), ~~~~~~ H~=~{\rm exp}(\beta u^{\widetilde{\beta}})
\nd
where $(\widetilde{\alpha},
\widetilde{\beta})=(3,3)$ and $(\alpha,\beta)=(0.1, 0.05)$. This choice is consistent with our ansatz $(\ref{metric1}),(\ref{metric2})$  and by
arranging the $D5-\bar{D5}-D7$ in similar fashion as in Fig 1, the warp factor in (\ref{warpy}) can be obtained.

\begin{figure}[htb]\label{dVSumax-1}
		\begin{center}
\includegraphics[width=0.6\textwidth,height=10cm,angle=-90]{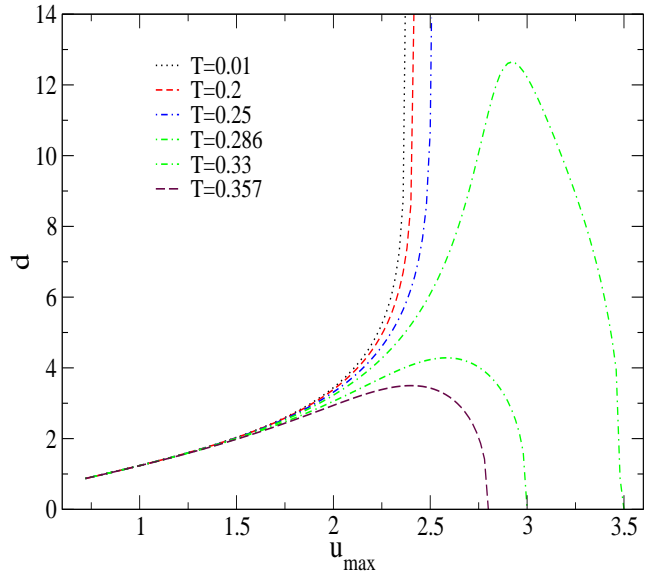}
		\caption{{Inter quark distance as a function of $u_{\rm max}$ for various temperatures and warp factor with
		$(\alpha,\widetilde{\alpha},\beta,\widetilde{\beta})=(0.1,3,0.05,3)$ in the warp factor equation.}}
		\end{center}
		\end{figure}
		\begin{figure}[htb]\label{dmaxvsT}
		\begin{center}
\includegraphics[width=0.6\textwidth,height=10cm,angle=-90]{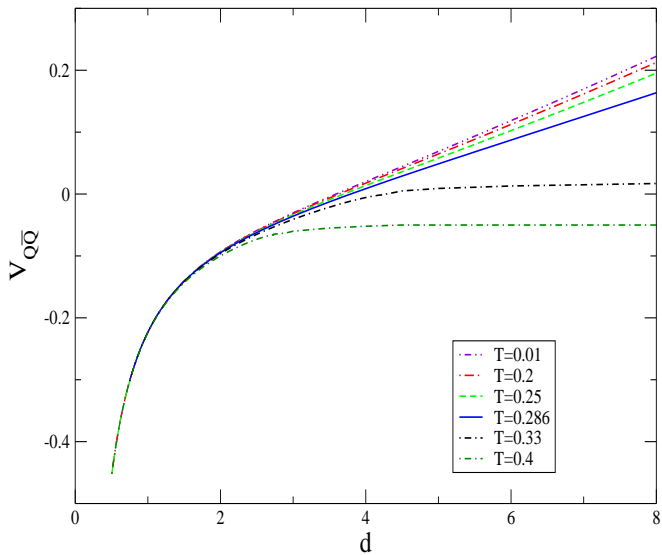}
		\caption{{Heavy quark potential $V_{Q\bar{Q}}$ as a function of quark separation $d$ with cubic warp
factor, or equivalently,
		$(\alpha,\widetilde{\alpha},\beta,\widetilde{\beta})=(0.1,3,0.05,3)$ in
the warp factor equation for various temperatures.}}
		\end{center}
		\end{figure}
		
	Observe that for a wide range of temperatures $0< 1/u_h\equiv T<1/u_c\equiv T_c$, the
potential and thus the free energy hardly changes. But near a narrow range of temperatures $T_{c} - \epsilon
< T < T_{c} + \epsilon$ (where $\epsilon \sim 0.05$),
free energy changes significantly .Our numerical analysis suggest the presence of a deconfinement transition, where
for a narrow range of temperatures $0.28 \le T_c  \le 0.39$ the free energy of
$Q\bar{Q}$ pair shows a sharp decline. Interestingly, changing the powers of $u$ in the exponential
changes the range of $T_c$ only by a small amount. So effectively $T_c$ lies in the range $0.2 \le T_c \le 0.4$.
Putting back units, and defining the {\it boundary} temperature\footnote{See sec. (3.1) of \cite{FEP} for details.}
$ \bar{T}$ as
$\bar{T} \equiv {g'(u_h)\over 4\pi\sqrt{h(u_h)}}$,
our analysis reveal:
\bg
\frac{0.91}{L^2} ~\le ~\bar{T}_c ~\le ~ \frac{1.06}{L^2}
\nd
which is the range of the melting temperatures in these class of theories for heavy quarkonium states. Since the
temperatures at both ends do not differ very much, this tells us that the melting temperature is inversely related to the
asymptotic AdS radius in large $N$ thermal QCD.

\section{Conclusions}
In this note we have summarized our proposal for the gravity dual of a non conformal finite temperature field theory
with matter in fundamental representation. The
gauge couplings run logarithmically in the IR while in the UV they become almost constant and the theory approaches
conformal fixed point. To our knowledge, the brane construction and the dual geometry presented here is the first
attempt to UV complete a Klebanov-Strassler type gauge theory with an asymptotically conformal field theory.

Although our
construction is rather technical with the gauge group being of the form $SU(N+M)\times SU(N)$ in the IR, one can perform
a cascade of Seiberg dualities to obtain the group $SU(\bar{M})$ and identify this with strongly coupled QCD. One may
even interpret that the gauge groups depicted in Fig 2 contain strongly coupled large N QCD. This is indeed
consistent  as the coupling  $g_{N+M}$ of $SU(N+M)$
factor  in
$SU(N+M)\times SU(N+M)$ or in $SU(N+M)\times SU(N)$,  {\it always} decreases as scale is increased. To see this, observe
that in the IR, the coupling $g_{N+M}$ runs logarithmically with scale and gets stronger as scale is decreased.
At the UV, we can arrange the
sources in the dual geometry such that $g_{N+M}$ decreases as the scale grows and runs as
$g_{M+N}\sim \widetilde{a}_k/\Lambda^k$. By
demanding\footnote{Note that $\widetilde{a}_k$ is different from $a_k$ used in 
(\ref{cond11}).}
\bg\label{dopm}
{k \widetilde{a}_k\over \Lambda^{k-1}} ~ > 0
\nd
we see that $g_{N+M}$ indeed decreases as scale is increased.
Thus from UV to IR
coupling always increases, just like QCD.

Of course the 't Hooft coupling
$\lambda_{N+M}=(N+M)g_{N+M}$ for the group
$SU(N+M)$ is still large in the limit $N\rightarrow \infty$, even if $g_{N+M}$ has decreased to very small value.  This allows us to use
the classical dual gravity description for the gauge theory which has a coupling that shrinks in the UV, mimicking  QCD. For even higher
energies, 't Hooft coupling will eventually become too small for finite $N+M$ and  supergravity description will no longer
hold.  But we can use perturbative
methods to analyze the field theory in the highest energies. For finite $N+M$ and very high energies,
the gauge coupling may even vanish giving rise to asymptotically free theory. These arguments lead us to conclude that,
in principle,
the brane configuration we proposed can incorporate QCD and
in the large N limit, the gravity dual we constructed can capture features of QCD.

Using the dual geometry, we have studied the dynamics of `quarks' in the gauge theory and computed shear viscosity $\eta$
and its
ratio to entropy $\eta/s$. The key to most of our analysis was the calculation of the stress tensor of gauge theory and
we showed how different
UV completions contribute to its expectation value. Using the correlation function for the stress tensor, we computed the shear viscosity of
the medium while the computation of pressure and energy density allowed us to calculate the entropy of the system.
Using a similar procedure to
calculate correlators of stress energy tensors, with introducing diagonal perturbations in the background metric,
we can easily evaluate the
bulk viscosity $\zeta$ of the non-conformal fluid. One can consider vector and tensor fields of higher rank to couple to the graviton
perturbations in the five dimensional effective theory and study how this coupling effects the bulk viscosity. The calculation is underway and
we hope to report on it in the near future.

All the calculations we performed regarding properties of the plasma did not account the effect of expansion of the medium
which is crucial in analyzing fluid dynamics. One possible improvement would be to construct a time dependent dual gravity which can describe
the expansion of the QGP formed in heavy ion collisions. A first attempt would be to consider collisions of open strings
ending on D7 branes and then compute their back-reactions on the geometry. The gravity waves associated with the collisions
evolve with time and from the induced boundary metric, one can compute the energy momentum tensor of the field theory. Analyzing the  time
dependence of this stress tensor, one can learn about the evolution of the medium and subsequently account for the effects
it has on the quark
dynamics.

We have computed the fluxes and the form of the warp factor for our static dual geometry, but did not give explicit
expressions for the deformation of the internal five dimensional metric to all order in $g_s N_f$. However, we have
explicitly shown the Einstein equations that determine the form of the internal metric and using our ansatz in \cite{jpsi1},
one can in principle compute coefficients in the expansion of the internal metric  to all orders in $g_sN_f$.
Even without a  precise knowledge of the
internal geometry, we were able to extract crucial information about the dual gauge theory and formulated how the higher order
corrections may enter into our analysis. Most of the calculation only relied on the warp factor and with the knowledge of its
precise form, we were able to calculate thermal mass, drag and diffusion coefficients, $\eta/s$ and finally free energy of
$Q\bar{Q}$ pair. For completeness of the supergravity analysis, we hope to compute the exact solution for the
internal metric using the ansatz of \cite{jpsi1} in our future work.

In our computation of the heavy quark potential, we classified the most general dual gravity that allows linear confinement of quarks at large
separation and small temperatures. We showed that if a gauge theory has dual gravity description and its effective degrees of freedom grows
 monotonically in the UV, it always shows linear confinement at large distances, as long as the dual warp factor satisfies a very simple relation given by
 (\ref{eq:finiteTopt}). Thus (\ref{eq:finiteTopt}) can be regarded as a sufficient condition for linear confinement of gauge theories with dual
 gravity. It would be interesting to study what are the general brane configurations that allow warp factors
 which satisfy (\ref{eq:finiteTopt}) and thus give rise to confining gauge theories.
 We leave it as a future direction to be explored.

As the dual geometry incorporates features of Seiberg duality cascade, our construction is ideal for studying phase transitions. The gauge
theories we studied have description in terms of gauge groups of lower and lower ranks. From the gravity dual analysis, by cutting the geometry
at certain radial location  and attaching another geometry up to infinity, we can construct gravity description for various phases of a gauge
theory. Each phase will have different dual geometries attached in the large $r$ region while the small $r$ region will be common to all the
theories. A flow from large $r$ to small $r$ geometry can be interpreted as `flow' from an effective theory in the UV to another in the IR.
From UV to IR, the different effective theories will describe different phases of the gauge theory and this can allow one to study the various
phases of dense matter. Thus our construction is not only useful to analyze strongly coupled gauge theory, but also has potential for studying
phase transitions in ultra dense medium and we hope to address this issue in the future.

\vskip.1in

\centerline{\bf Acknowledgements}

\vskip.1in

\noindent Our work was supported in part by the Natural Sciences and Engineering
Research Council of Canada, and in part by the Office of Nuclear Science of
the US Department of Energy under grant No. DE-FG02-93ER40764.

\end{document}